\RequirePackage{fix-cm}
\documentclass[twocolumn]{svjour3}          
\smartqed  
\usepackage{graphicx}
\usepackage{lineno,hyperref}
\usepackage{latexsym}
\usepackage{lipsum}
\usepackage{graphicx}
\usepackage{subcaption}
\usepackage{amsmath}
\usepackage{mathtools}
\usepackage{caption}
\usepackage{color, soul}
\usepackage{mathptmx}      

\journalname{Nonlinear Dynamics}

\begin{document}

\title{Simplified model and analysis of a pair of coupled thermo-optical MEMS oscillators}


\author{Richard H. Rand          \and
        Alan T. Zehnder \and B. Shayak \and Aditya Bhaskar 
}


\institute{Richard H. Rand          \and
        Alan T. Zehnder \and B. Shayak \and Aditya Bhaskar \at
              Theoretical and Applied Mechanics, Sibley School of Mechanical and Aerospace Engineering, Cornell University,
Ithaca, New York 14853 USA  \\
              \email{rhr2@cornell.edu}           
           \and
           Richard H. Rand \at
              Department of Mathematics, Cornell University, Ithaca, New York 14853 USA
}

\date{Received: date / Accepted: date}

\maketitle

\begin{abstract}
{Motivated by the dynamics of micro-scale oscillators with thermo-optical feedback, a simplified third order model, capturing the key features of these oscillators is developed, where each oscillator consists of a displacement variable coupled to a temperature variable. Further, the dynamics of a pair of such oscillators coupled via a linear spring is analyzed. The analytical procedures used are the variational equation method and the two-variable expansion method. It is shown that the analytical results are in agreement with the results of numerical integration. The bifurcation structure of the system is revealed through a bifurcation diagram.}
\end{abstract}

\section{Introduction}
\label{intro}

The dynamics of resonant MEMS, or micro-electro mechanical systems, are of interest due to both the large number of current and potential applications \cite{Younis} and for the new questions that arise in their design and analysis. Of interest here are MEMS systems consisting of a resonator suspended above a substrate and illuminated with a focused laser.  The suspended resonator and substrate form a Fabry-P\'{e}rot interferometer \cite{group1} with the net result that the amount of laser heating of the resonator is modulated by the resonator/substrate gap.  The resulting feedback of heating, temperature and thermal stress can result in limit cycle oscillations \cite{group2}.  Prior work has shown that these oscillators can be entrained \cite{Pandey1} to inertial and optical signals \cite{Pandey2}.

Accounting for laser heating and thermo-mechanical coupling the oscillator can be described by the
following equation of motion \cite{group4}
\begin{subequations} \label{fullgoverning}
\begin{equation} \label{dispeqn}
\ddot{z}+\frac{1}{Q}\dot{z}+\left( 1+CT \right)z+\beta {{z}^{3}}=DT
\end{equation}
\begin{equation} \label{Teqn}
\dot{T}=H{{P}_{L}}\left[ \alpha +\gamma {{\sin }^{2}}\left( 2\pi (z-{{z}_{0}}) \right) \right]-BT
\end{equation}
\end{subequations}
where $z$ is the displacement of the oscillator from its equilibrium position and $T$ is the temperature of the oscillator. In Eq.~\eqref{dispeqn}, \(z\) is the out-of-plane displacement of the oscillator, normalized by the wavelength of the laser light, \(Q\) is the quality factor, \(C\) is the thermal coefficient for linear stiffness, \(\beta\) is the cubic stiffness, \(D\) is the static position per unit change in temperature, \(E\) is the inertial drive amplitude, and \(\omega\) is the inertial drive frequency. In Eq.~\eqref{Teqn}, \(T\) is the average temperature of the oscillator, \(B\) and \(H\) are thermal constants,  \(P_L\) is the continuous-wave laser power, \(\alpha\) is the minimum absorption, \(\gamma\) is the contrast in absorption, and \(z_0\) is the equilibrium position of the oscillator with respect to the absorption curve. In Eq.~\eqref{dispeqn}, the linear stiffness and the thermal driving force depend on the temperature and in Eq.~\eqref{Teqn}, the laser absorption is modulated by the displacement of the oscillator \cite{group3}.

The nonlinearities and presence of a large number of parameters make the equations intractable and thus we develop a model that has the essential features of the system but is amenable to analysis. The premise of our study is that understanding the dynamics of the simplified model will help in explaining phenomena associated with the more complex model, and the corresponding experiments.  The dynamics of a single oscillator are first discussed followed by an analysis of the dynamics of a coupled pair of oscillators.  

To the best of one's knowledge this is one of the first studies where coupled third order oscillators have been considered. The archetypical system for synchronization is the phase only oscillator, where each oscillator has the structure $\dot{\theta} = \omega +f(\theta, t)$. Here $\theta$ denotes the phase of each oscillator while $f(\theta, t)$ represents various forms of coupling and/or nonlinear term. This kind of study was pioneered by Kuramoto \cite{Kuramoto} and those results were later applied to biological situations by Mirollo and Strogatz \cite{Strogatz}. The case of two coupled van der Pol oscillators was first studied by Storti and Rand \cite{Storti}. They obtained the regions of in-phase and out-of-phase locking as well as drifting. 

In recent years, interest has arisen in a multitude of other kinds of oscillators. Valente et. al. \cite{Valente} have considered a system consisting of two masses coupled via a spring, which can rigidly impact a fixed stop. Hybrid periodic orbits are found as well as trajectories where there are infinitely many impacts with the stop occurring in a finite time. Fradkov and Andrevsky \cite{Fradkov} have studied synchronization of two simple pendula coupled by a linear spring. Sliwa and Grygiel \cite{Sliwa} have considered a pair of coupled Kerr oscillators. These have the equation of the structure $dz/dt = -j\omega z-j \epsilon a*a^{2}+Fe^{-j \Omega t}-\gamma z$ where $z$ is a complex variable, $j$ denotes the imaginary unit and all other symbols denote constants. This oscillator arises in nonlinear optics. Switching of the system between different semi-stable attractors as well as chaotic beats have been observed. Suchorsky and Rand \cite{Suchorsky} have considered van der Pol oscillators coupled by fractional derivative. The authors have considered regions of locking and drifting and have demonstrated the reduction to known results in the limits where the fractional derivative is replaced by an integer. Finally, Chavez et. al. \cite{Chavez} have considered a pair of forced Duffing oscillators coupled via soft impacts. A bifurcation diagram has been presented by these authors. 

The previous paragraphs give some indication of the breadth and variety of oscillators and coupling mechanisms which have been considered in literature. However, all of these either consider phase-only or second order models. The introduction of a higher order system, in this work, presents new and interesting phenomena which will be discussed. The complexity of the system requires the use of a multitude of techniques. We will use a variational equation method, a regular perturbation theory on a Mathieu-like equation as well as a two-variable expansion on the system equations to generate a set of slow flow equations.

\section{Simplified model of one oscillator} \label{section2}

A key observation from the analysis of Eq.~\eqref{fullgoverning}, is the presence of a limit cycle for a large enough value of $P_L$ \cite{group1}. Hence to emulate the full equations the reduced model must also support a limit cycle. The damping in Eq.~\eqref{dispeqn} is inessential as there is already a damping term in Eq.~\eqref{Teqn}, and that alone provides all the damping which we require for off-cycle motions to die out. The cubic nonlinearity term and temperature coefficient in the linear stiffness term in Eq.~\eqref{dispeqn} are not essential for the formation of a limit cycle. In Eq.~\eqref{Teqn}, $\sin ^{2}(z-z_{0})$ can be Taylor expanded as \\${{\left[ \left( z-{{z}_{0}} \right)-{{\left( z-{{z}_{0}} \right)}^{3}}/6+... \right]}^{2}}$. Neglecting terms of order $O((z-z_0)^4)$ and higher, this term can be further simplified to $z^{2}-zz_{0}$ without losing any essential features. Thus Eq.~\eqref{fullgoverning} can be reduced to the following:

\begin{subequations} \label{simplified}
\begin{equation} \label{dispeqn2}
\ddot{z}+z=T
\end{equation}
\begin{equation} \label{Teqn2}
\dot{T}+T={{z}^{2}}-z{{z}_{0}}
\end{equation}
\end{subequations}

A Lindstedt Poincar\'{e} type analysis  \cite{Rand} of this system is performed, assuming the amplitude of oscillation to be small, of order $\varepsilon$. Further, the assumption is made that the parameter $z_0$ is of the second order of smallness, and it is scaled as $z_0 = \varepsilon ^{2} p$ to get 
\begin{subequations} \label{eqn3}
\begin{equation} \label{dispeqn3}
\ddot{z}+z=T 
\end{equation}
\begin{equation} \label{Teqn3}
\dot{T}+T=\varepsilon {{z}^{2}}-{{\varepsilon }^{2}}pz 
\end{equation}
\end{subequations}
Using the perturbation theory, we expand $z=z_{(0)}+\varepsilon z_{(1)}+\varepsilon ^{2} z_{(2)}$, $T=T_{(0)}+\varepsilon T_{(1)}+\varepsilon^{2}T_{(2)}$ and use the time dilation, $\tau = \omega t$, where $ \omega = 1+k_{1} \varepsilon +k_{2} \varepsilon ^{2}$. At the lowest order we get,
\begin{subequations}
\begin{equation} \label{dispeqn4}
{{z}_{(0)}}''+{{z}_{(0)}}={{T}_{(0)}} 
\end{equation}
\begin{equation} \label{Teqn4}
{{T}_{(0)}}'+{{T}_{(0)}}=0 
\end{equation}
\end{subequations}
where prime denotes differentiation relative to the stretched time, $\tau$. Since Eq.~\eqref{Teqn4} has exponentially decaying solutions, they are not of interest in driving a limit cycle so we take $T_{(0)}=0$. With this substitution, Eq.~\eqref{dispeqn4} has the standard oscillatory solutions. Since the system is autonomous, the phase is arbitrary and the oscillation can be treated as $z_{(0)}=A\cos \tau $. Balancing the coefficients of $\varepsilon$ we get,
\begin{subequations}
\begin{equation} \label{dispeqn5}
{{z}_{(1)}}''+{{z}_{(1)}}=-2{{k}_{1}}{{z}_{(0)}}''+{{T}_{(1)}}
\end{equation}
\begin{equation} \label{Teqn5}
{{T}_{(1)}}'+{{T}_{1}}=z_{(0)}^{2}-p{{z}_{(0)}}-{{k}_{1}}{{T}_{(0)}}' 
\end{equation}
\end{subequations}
In Eq.~\eqref{Teqn5}, since $T_{(0)}=0=T'_{(0)}$, the last term drops out. Removal of resonance terms from Eq.~\eqref{dispeqn5} gives $k_{1}=0$ but yields no information about $A$, which carries along as a parameter, giving $z_{(1)}$ and $T_{(1)}$ in terms of $A$. Balancing the coefficients of $\varepsilon^2$ in the perturbation analysis we get,
\begin{subequations}
\begin{equation} \label{dispeqn6}
{{z}_{(2)}}''+{{z}_{(2)}}-2{{k}_{1}}{{z}_{(1)}}''+\left( -2{{k}_{2}}-k_{1}^{2} \right){{z}_{(0)}}''+{{T}_{(2)}}
\end{equation}
\begin{equation} \label{Teqn6}
{T_{(2)}}''+T_{(2)}=(2z_{(2)}-p)z_{(1)}-k_{1}{T_{(2)}}'-k_{2}{T_{(0)}}'
\end{equation}
\end{subequations}
Removing secular terms from Eq.~\eqref{dispeqn6} yields the following solution:
\begin{subequations}\label{solution7}
\begin{equation} \label{dispeqn7}
z=A \cos\omega t+ {{A}^{2}}\left( \frac{1}{2}-\frac{1}{15}\sin 2\omega t-\frac{1}{30}\cos 2\omega t \right) 
\end{equation}
\begin{equation} \label{Teqn7}
T={{A}^{2}}\left( \frac{1}{2}+\frac{1}{5}\sin 2\omega t+\frac{1}{10}\cos 2\omega t \right)
\end{equation}
\end{subequations}
where 
\begin{subequations}
\begin{equation} \label{Aeqn}
A=\sqrt{10p} /3
\end{equation}
\begin{equation} \label{omegaeqn}
\omega = 1-p/27 
\end{equation}
\end{subequations}
Since the perturbation theory is developed in terms of the amplitude, one is not required to explicitly report $\varepsilon$  in these expressions. It is noted that $A$ and $\omega$ are the amplitude and frequency of the fundamental motion $z_{(0)}$, respectively. Thus Eq.~\eqref{simplified} exhibits limit cycle oscillations for all positive values of $p$. The amplitude increases and frequency decreases with increasing $p$.

\section{Coupled pair of oscillators and variational equation method}

We now turn to the analysis of two coupled MEMS oscillators, here assumed to be identical. The case where detuning is present will be considered in a later study. Linear coupling is assumed, which could be achieved either via a spring or via electrostatic coupling. The equations of the coupled oscillators are
\begin{subequations} \label{coupled}
\begin{equation} \label{Cdispeqn1}
{{{\ddot{z}}}_{1}}+{{z}_{1}}={{T}_{1}}+\alpha \left( {{z}_{2}}-{{z}_{1}} \right) 
\end{equation}
\begin{equation} \label{CTeqn1}
 {{{\dot{T}}}_{1}}+{{T}_{1}}=z_{1}^{2}-{{z}_{1}}p
\end{equation}
\begin{equation} \label{Cdispeqn2}
 {{{\ddot{z}}}_{2}}+{{z}_{2}}={{T}_{2}}+\alpha \left( {{z}_{1}}-{{z}_{2}} \right) 
 \end{equation}
\begin{equation} \label{CTeqn2}
{{{\dot{T}}}_{2}}+{{T}_{2}}=z_{2}^{2}-{{z}_{2}}p 
\end{equation}
\end{subequations}
where $\alpha$ is the coupling stiffness. The model consists of two identical coupled oscillators, and hence in-phase (IP) and out of phase (OP) modes are expected. Here , the IP mode is defined as $z_1(t) = z_2(t)$ and $T_1(t)=T_2(t)$ whereas the OP mode is defined as $z_1(t) = -z_2(t)$ and $T_1(t)=-T_2(t)$. Indeed, linearization about the origin yields all six eigenvectors to have the structure corresponding to IP and OP modes. Two of the modes are associated with negative real eigenvalues for all $p$ and $\alpha$. The four remaining eigenvalues are two complex conjugate pairs, which are purely imaginary for $p=0$ and have positive real parts when $p>0$. One pair of these eigenvalues attaches to a pair of eigenvectors which has the IP mode shape while the other pair attaches to an eigenvector pair which has the OP mode shape. It is noted that in the full nonlinear system, the change in sign of the eigenvalues as $p$ is increased through zero corresponds to the Hopf bifurcation which gives rise to limit cycle oscillations. 

Eq.~\eqref{coupled} is transformed in the following way : 
\begin{subequations} \label{transform}
\begin{equation} \label{disptransform1}
x={{z}_{1}}+{{z}_{2}} 
\end{equation}
\begin{equation} \label{disptransform2}
y={{z}_{1}}-{{z}_{2}}
\end{equation}
\begin{equation} \label{Ttransform1} 
u={{T}_{1}}+{{T}_{2}} 
\end{equation}
\begin{equation} \label{Ttransform2}
v={{T}_{1}}-{{T}_{2}} 
\end{equation}
\end{subequations}
Substituting Eq.~\eqref{transform} into Eq.~\eqref{coupled}, one gets,
\begin{subequations} \label{Cmanip}
\begin{equation} \label{Cmanip1}
\ddot{x}+x=u
\end{equation}
\begin{equation} \label{Cmanip2}
\dot{u}+u=\frac{{{x}^{2}}+{{y}^{2}}}{2}-px 
\end{equation}
\begin{equation} \label{Cmanip3}
\ddot{y}+\left( 1+2\alpha  \right)y=v
\end{equation}
\begin{equation} \label{Cmanip4}
\dot{v}+v=xy-py
\end{equation}
\end{subequations}
The presence of IP mode is clearly demonstrated by these equations - in this mode $y=v=0$, and $x$ and $y$ satisfy Eq.~\eqref{simplified}. An exact OP mode however does not exist; setting $x=u=0$ leads to an apparent contradiction. The problem of persistence of linearized modes in nonlinear systems has been considered by Hennig \cite{Hennig}. \\

\subsection{Stability of the IP mode}
The stability of the IP mode is now analyzed by constructing the variational equation, as in Stoker \cite{Stoker}. Small perturbations $\delta x$, $\delta y$, $\delta u$ and $\delta v$ are introduced on top of the steady state solutions of $x$,$y$,$u$ and $v$, where the steady state solutions are characterized by $x = 2z$ and $u=2T$, where $z$ and $T$ are given by Eq.~\eqref{solution7}, and $y$ and $v$ are zero. This variational equation for the $\delta y$-$\delta v$ system can be combined into a single third order equation as
\begin{equation} \label{combinedthirdorder}
\delta \dddot{y}+\delta \ddot{y}+\left( 1+2\alpha  \right)\delta \dot{y}+\left( 1+2\alpha +p-F(t) \right) \delta y=0
\end{equation}
where $F(t)$ is the limit cycle motion exhibited by $x(t)$ whose stability is being sought. Since $x=z_1+z_2$ which equals $2z_1$ for the IP mode, and since the limit cycle for a single oscillator was found in Eq.~\eqref{solution7}, one may write 
\begin{equation} \label{Feq2z}
F(t)=2A \cos\omega t+ 2{{A}^{2}}\left( \frac{1}{2}-\frac{1}{15}\sin 2\omega t-\frac{1}{30}\cos 2\omega t \right). 
\end{equation}

\begin{figure}
	\centering
		\includegraphics[width=0.5\textwidth]{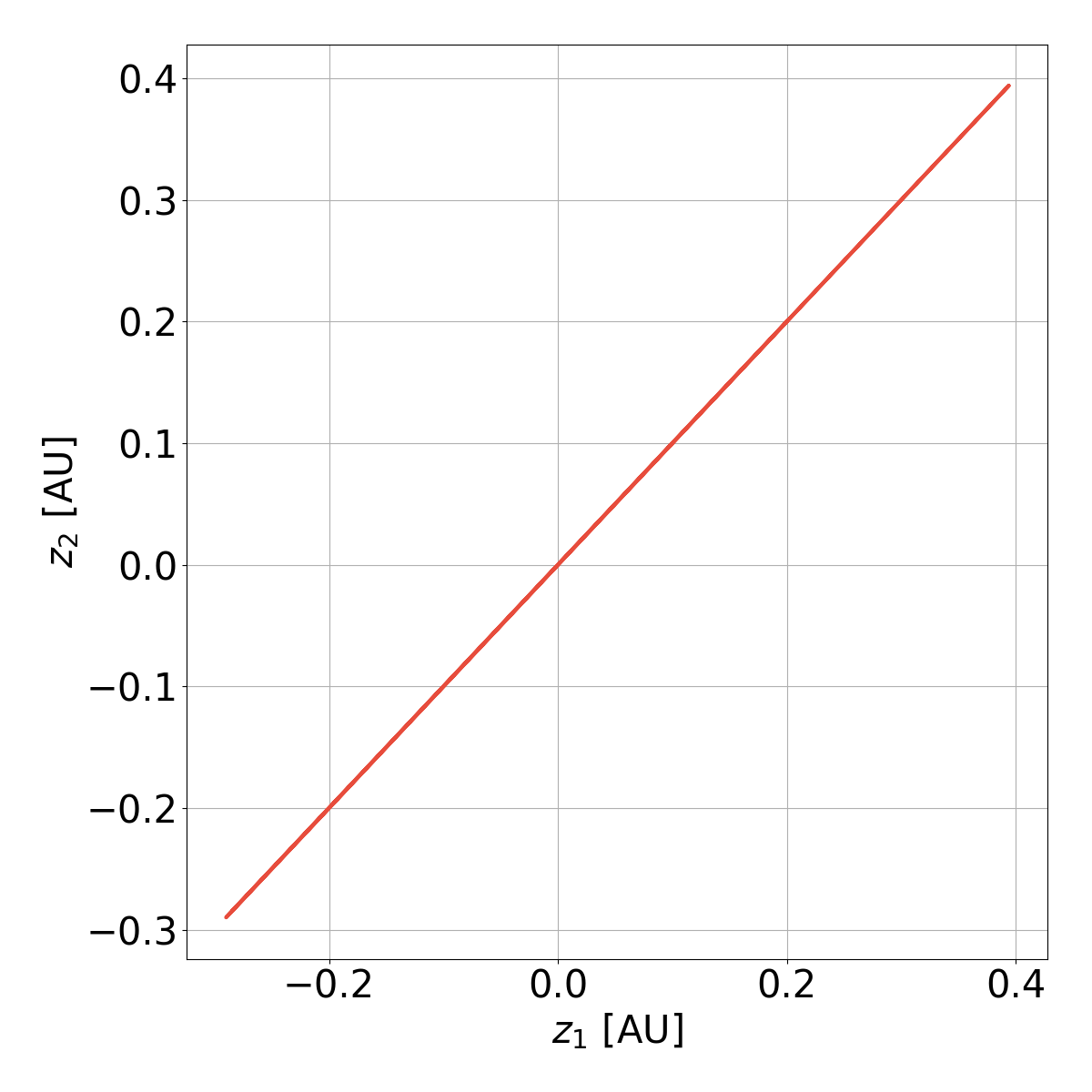}
	\caption{Phase plane showing the in-phase mode. Steady state has been shown after the initial transients decay. This plot is based on the numerical integration of the original system equations.}	
	\label{fig:IPmode}
\end{figure}

\begin{figure}
	\centering
	\includegraphics[width=0.5\textwidth]{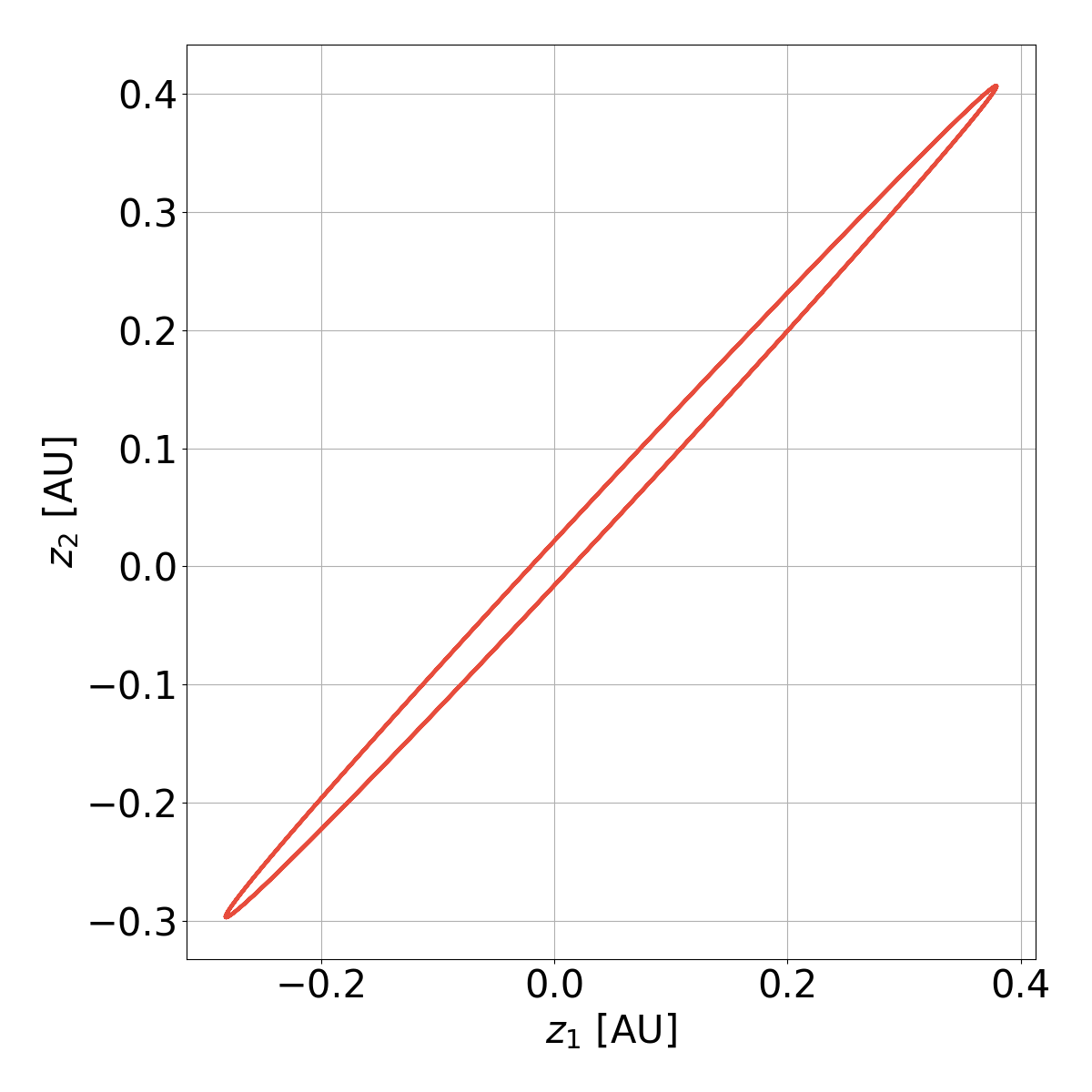}
	\caption{Branching out of the in-phase mode into an ellipse at $\alpha = 0.055$. Steady state has been shown after the initial transients decay. This plot is based on the numerical integration of the original system equations.}
	\label{fig: Ellipse}
\end{figure}

\begin{figure}
	\centering
	\includegraphics[width=0.5\textwidth]{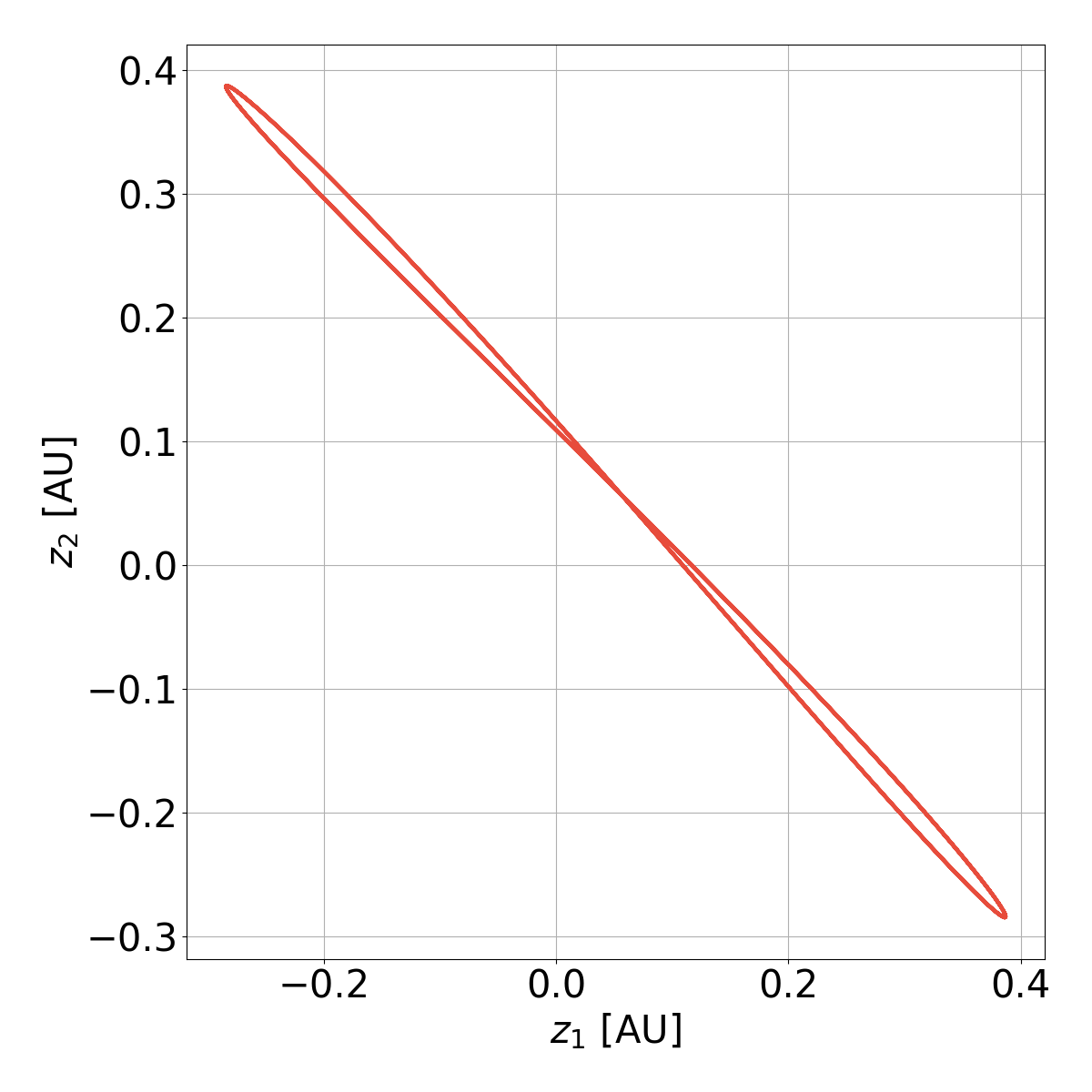}
	\caption{Phase plane showing an elongated figure-of-eight OP-like mode. This plot is based on the numerical integration of the original system equations.}
	\label{fig: OPmode}
\end{figure}

A perturbative approach will now be used to determine the stability of the IP mode. For this approach it is sufficient to retain only the first term in the right hand side of Eq.~\eqref{Feq2z}. From this point onwards, $\delta y$ is written as $q$ for notational convenience. To carry out a perturbation procedure the variables are rescaled as follows :
\begin{equation} \label{combinedthirdorder2}
\omega ^3 q'''+\omega ^2 q''+\nu \omega q'+\left( \nu +\mu \cos \tau +\mu^2 p \right)q=0.
\end{equation}
where $\tau = \omega t$, the prime denotes differentiation with respect to the variable $\tau$, $\mu$ is the amplitude of the parametric excitation and $\nu=1+2\alpha$. We use the expansion, $ \nu = 1+ \mu \nu_{1} + \mu^{2} \nu_{2}$, let $\omega = 1-\mu^2p/27$ and then expand $q=q_{0}+\mu q_{1}+\mu^{2}q_{2}$. Before proceeding further, an outline of the philosophy behind the perturbation is given. To obtain the transition curves of  Eq.~\eqref{combinedthirdorder}, one seeks the motion that is periodic on the curve itself, with a period equal to that of $F(t)$. The choice of period is motivated by the fact that in the absence of the perturbations, Eq.~\eqref{combinedthirdorder2} possesses oscillatory solutions with natural frequency 1; since the frequency of the parametric excitation is also perturbatively close to 1, we are close to a 1:1 resonance of the system. Substituting the perturbation expansions into Eq.~\eqref{combinedthirdorder2}, at zero order one has
\begin{equation} \label{zeroorder}
q_{0}'''+q_{0}''+q_{0}'+q_{0}=0  
\end{equation} 
which has the fundamental solutions $\cos \tau$, $\sin \tau$ and $\operatorname{e}^{-\tau}$. The exponentially decaying solution is not of interest so we let 
\begin{equation} \label{Linearcombo}
q_{0} = m\cos \tau+n\sin \tau 
\end{equation}
where $m$ and $n$ are arbitrary. Equating terms of order $\mu$ in the expanded form of Eq.~\eqref{combinedthirdorder2}, and substituting Eq.~\eqref{Linearcombo} into the result, we get
\begin{equation} \label{firstorder}
q_1'''+q_1''+q_1'+q_1=-\frac{m}{2} \left(1+\cos 2\tau \right)-\frac{n}{2}\sin 2\tau .
\end{equation}
The resonant terms can be removed from this equation if one sets $\nu_1 = 0$. Using the method of undetermined coefficients, one finds
\begin{equation} \label{firstordersoln}
q_{1}=\frac{(m-2n)\cos 2\tau}{30}+\frac{(2m+n)\sin 2\tau}{30}-\frac{m}{2} 
\end{equation}
At order $\mu^2$ we get
\begin{subequations}\label{secondorder}
\begin{equation*} 
q_2'''+q_2''+q_2'+q_2 = -q_{1}\cos \tau -(p/9)q_0'''+(2/27)p q_0''
\end{equation*}
\begin{equation}
-(\nu_{2}-(p/27))q_0'-(p+\nu_{2})q_{0}
\end{equation}
\end{subequations}
where $q_{0}$ and $q_{1}$ have been determined above. This contains resonance terms on the right hand side; removal of which requires 
\begin{subequations} \label{Resonsant}
\begin{equation} \label{Resonsant1}
\frac{2np}{27}+\frac{29mp}{27}+\nu_{2} n-\frac{n}{30}+\nu_{2} m -\frac{29m}{60} = 0 
\end{equation}
\begin{equation} \label{Resonsant2}
\frac{29np}{27}-\frac{2mp}{27}+\nu_{2} n+\frac{n}{60}-\nu_{2} m +\frac{m}{60} = 0 
\end{equation}
\end{subequations}
Equations \eqref{Resonsant} are a pair of simultaneous linear homogeneous equations; a non-trivial solution exists if and only if the determinant of the matrix vanishes. This condition leads to
\begin{equation} \label{detzero}
\frac{67600p^2+133920 \nu_{2}p-29520p+116640\nu_{2}^{2}-31104\nu_{2}-405}{58320} =0 
\end{equation}
For $p=0.1$ the value of $\nu_{2}$ obtained from Eq.~\eqref{detzero} is 0.2455. Now, $\nu$ is $1+\mu^{2} \nu_{2}$ which also equals $1+2\alpha$. Since $\mu$ is twice the amplitude of the single oscillator limit cycle, determined earlier to be 0.333 using Eq.~\eqref{Aeqn}, the transitional value of $\alpha$, above which the IP mode is stable and below unstable, is found to be 0.0545. 
The theoretical predictions are compared with the numerical simulations. At $\alpha = 0.1$, with initial conditions $\{z_1(0), z_2(0), \dot{z}_1(0), \dot{z}_2(0), T_1(0), T_2(0)\} = \{0.1, 0.09, 0, 0, 0, 0\}$, close to the IP mode, the numerical solution shows an IP mode as shown in Fig. \ref{fig:IPmode}. For the same initial conditions, as $\alpha$ is decreased to 0.055, the straight line of the IP mode branches out into an ellipse as show in Fig \ref{fig: Ellipse}. The critical value of $\alpha$ is very close to the value predicted by the variational method. For initial conditions, $\{z_1(0), z_2(0), \dot{z}_1(0), \dot{z}_2(0), T_1(0), T_2(0)\} = \{0.1, -0.09, 0, 0, 0, 0\}$, close to the OP mode, the numerical solution resembles an OP motion as shown in Fig. \ref{fig: OPmode}. Although the IP mode is an exact straight line, the OP mode is not.

An intriguing feature is that the variational analysis on the $x$,$u$ equations, considering perturbations $\delta x$ and $\delta u$ off their steady state values, leads to the same equation as Eq.~\eqref{combinedthirdorder} but with $\alpha =0$. This is supposed to be unstable. However, we recall that a variational analysis gives \emph{Lyapunov} stability whereas the limit cycles possesses only \emph{orbital} stability\cite{Stoker}. In the same way, a variational analysis of the limit cycle of van der Pol oscillator also yields a ‘spurious’ instability. The source of the instability is phase shear. As  the trajectories approach the limit cycle, their period change, reaching a limiting value at the cycle itself. If the motion of a point exactly on the limit cycle is compared with that of a point, $\varepsilon$ distance away from it, then it will be seen that the the separation between the two points initially increases from $\varepsilon$ to a finite value due to the unequal periods of their trajectories.

\subsection{Stability of the OP mode}

It is now desired to determine the stability of the OP-like mode. It was mentioned earlier that the strict OP mode $z_1=-z_2$ and $T_1=-T_2$ does not exist for Eqs.~\eqref{coupled}. In order to investigate the stability of the motion similar to the OP mode (cf. Fig.~\ref{fig: OPmode}), one first needs to obtain an expression for the motion, which one will do using LINDSTEDT's method. The variable scaling is the same as that of a single oscillator Eq.~\eqref{eqn3}, and the solutions $z_{1}=A\cos(\omega t)$ and $z_{2}=-A\cos (\omega t)$ are imposed at the $\varepsilon ^{0}$ level. Computer algebra is used for the calculation, owing to intractability of the mathematical manipulations. As in Section I, the starting step is to introduce the parameter $\varepsilon$, in a manner consistent with the scalings for a single oscillator Eq.~\eqref{eqn3}. Time is rescaled as $\tau =(1+k_{1} \varepsilon +k_{2}\varepsilon^{2})t$ and the variables are expanded as $z_{1}=z_{10}+\varepsilon z_{11}+\varepsilon ^{2}z_{12}$, and so on. The zero order equations are
\begin{subequations} \label{Lind}
\begin{equation} \label{Lind1}
{{z}_{10}}''+\left( 1+\alpha  \right){{z}_{10}}=\alpha {{z}_{20}}+{{T}_{10}}
\end{equation}
\begin{equation} \label{Lind2}
{{T}_{10}}'+{{T}_{10}}=0 
\end{equation}
\begin{equation} \label{Lind3}
{{z}_{20}}''+\left( 1+\alpha  \right){{z}_{20}}=\alpha {{z}_{10}}+{{T}_{20}} 
\end{equation}
\begin{equation} \label{Lind4}
{{T}_{20}}'+{{T}_{20}}=0 
\end{equation}
\end{subequations}

It is found that the $z$'s are oscillatory and the $T$'s are damped. Neglecting the exponentially decaying terms, the frequency of oscillation of Eqs. \eqref{Lind1} and \eqref{Lind3} is obtained as $(1+2\alpha )^{1/2}$. Now, we impose externally that $z_{10}=A\cos \omega \tau$ and $z_{20}=-A\cos \omega \tau$. The order $\varepsilon$ equations are
\begin{subequations} \label{ordere}
\begin{equation} \label{ordere1}
{{z}_{11}}''+\left( 1+\alpha  \right){{z}_{11}}={{T}_{11}}+\alpha {{z}_{21}} 
\end{equation}
\begin{equation} \label{ordere2}
{{T}_{11}}''+{{T}_{11}}=z_{10}^{2} 
\end{equation}
\begin{equation} \label{ordere3}
{{z}_{21}}''+\left( 1+\alpha  \right){{z}_{21}}={{T}_{21}}+\alpha {{z}_{11}}
\end{equation}
\begin{equation} \label{ordere4}
{{T}_{21}}''+{{T}_{21}}=z_{20}^{2}
\end{equation}
\end{subequations}

Plugging the zero order solutions into these equations leads to a coupled inhomogeneous system of ODEs for the 1-level variables. To uncouple this system the transformation to sum and difference coordinates is performed with $z_{11}=(u+v)/2$ and $z_{21}=(u-v)/2$. Removal of the resonance term gives $k_{1}=0$. Solving the system in the new variables and inverting gives $z_{11}$ and $z_{21}$ as trigonometric functions of $t$, with the amplitude $A$ still as a parameter. 

Determination of amplitude is achieved from the next level. The equations at this level are
\begin{subequations} \label{ordere2}
\begin{equation} \label{ordere21}
{{z}_{12}}''+\left( 1+\alpha  \right){{z}_{12}}=\alpha {{z}_{22}}+{{T}_{12}}+2{{k}_{2}}{{z}_{10}}'' 
\end{equation}
\begin{equation} \label{ordere22}
{{T}_{12}}'+{{T}_{12}}={{k}_{2}}{{T}_{10}}'+\left( p-2{{z}_{11}} \right){{z}_{10}}
\end{equation}
\begin{equation} \label{ordere23}
{{z}_{22}}''+\left( 1+\alpha  \right){{z}_{22}}=\alpha {{z}_{12}}+{{T}_{22}}+2{{k}_{2}}{{z}_{20}}'' 
\end{equation}
\begin{equation} \label{ordere24}
{{T}_{22}}'+{{T}_{22}}={{k}_{2}}{{T}_{20}}'+\left( p-2{{z}_{21}} \right){{z}_{20}} 
\end{equation}
\end{subequations}
Substitution of the already determined 0- and 1- level quantities into Eq.~\eqref{ordere2}, followed by yet another sum and difference transformation, leads to a situation where the resonance terms can be removed. Computer algebraic manipulations yield the following :
\begin{subequations} \label{Lindz}
\begin{equation} \label{Lindz1}
{{z}_{1}}=A\cos \omega t+\varepsilon u /2
\end{equation}
\begin{equation} \label{Lindz2}
{{z}_{2}}=-A\cos \omega t+\varepsilon u /2
\end{equation}
\end{subequations}
where
\begin{subequations} \label{omegaAu}
\begin{equation} \label{omegaAu1}
\omega =\sqrt{1+2\alpha} 
\end{equation}
\begin{equation} \label{omegaAu2}
A=\frac{2p\left( 8\alpha +3 \right)\left( 8\alpha +5 \right)}{{{\left( 128{{\alpha }^{2}}+128\alpha +27 \right)}^{1/2}}}
\end{equation}
\begin{equation} \label{omegaAu3}
u={{A}^{2}}-\frac{2\omega \sin 2\omega t+\cos 2\omega t}{\left( 8\alpha +3 \right)\left( 8\alpha +5 \right)} 	
\end{equation}
\end{subequations}
\begin{figure}
	\centering	
	\includegraphics[width=0.5\textwidth]{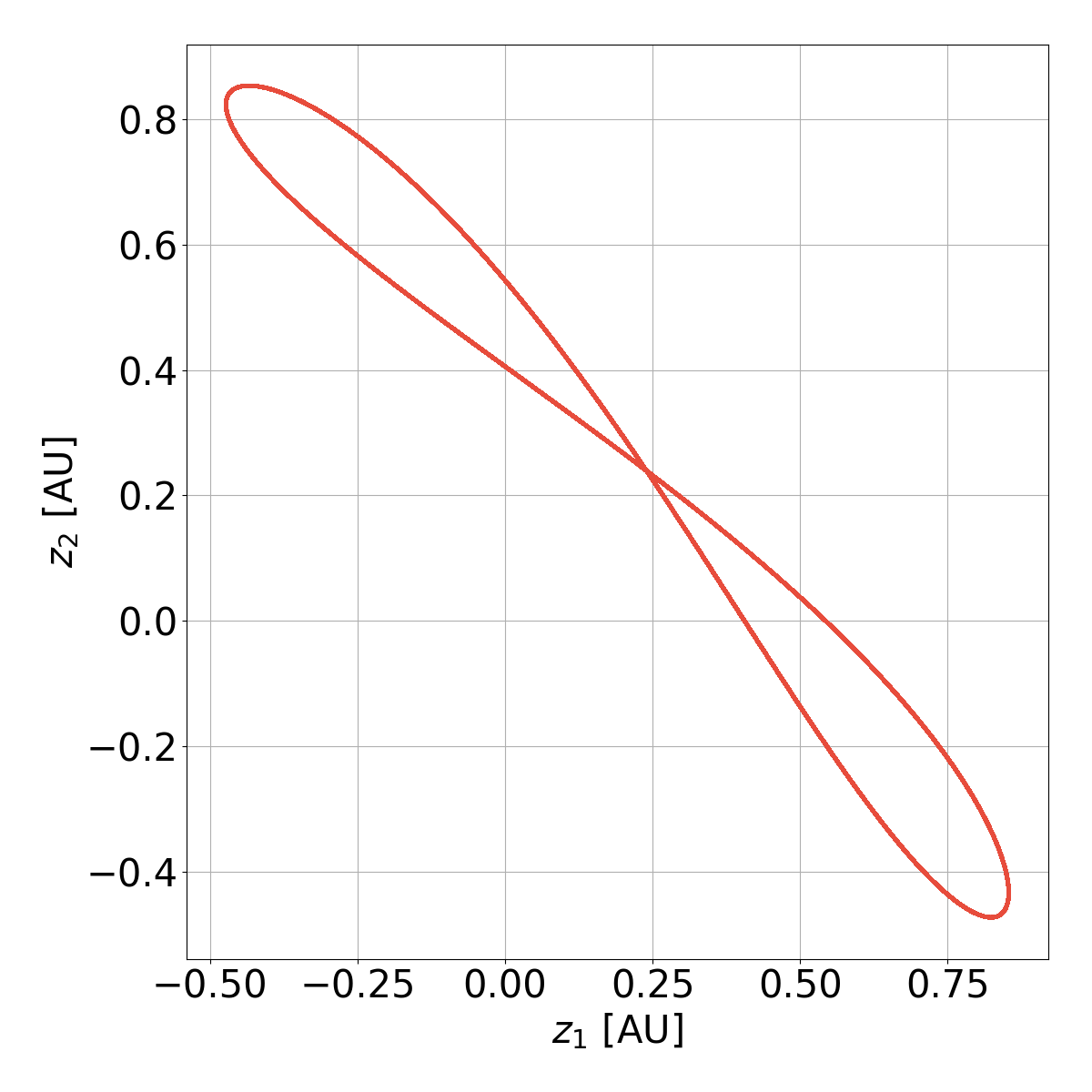}
	\caption{Phase portrait of the OP mode from perturbation analysis}
	\label{fig: OPPerturb}
\end{figure}
The trajectory predicted by Eqs.~\eqref{Lindz} and\eqref{omegaAu} is plotted in in Fig.~\ref{fig: OPPerturb} as a phase portrait, which has the desired figure-of-eight shape but is wider than the actual, as in Fig. \ref{fig: OPmode}. This solution will now be substituted into Eqs.~\eqref{coupled} to construct the linear variational equation for perturbations $\delta z_1$, $\delta z_2$, $\delta T_1$ and $\delta T_2$ added onto the ``steady state'' solutions $z_1^{\ast}(t)$, $z_2^{\ast}(t)$, $T_1^{\ast}(t)$ and $T_2^{\ast}(t)$. The linear variational equation turns out to be
\begin{subequations} \label{6var}
\begin{equation} \label{6var1}
{{{\ddot{\delta z}}}_{1}}+{{\delta z}_{1}}={{\delta T}_{1}}+\alpha \left( {{\delta z}_{2}}-{{z}_{1}} \right) 
\end{equation}
\begin{equation} \label{6var2}
{{{\dot{\delta T}}}_{1}}+{{\delta T}_{1}}=2z_1^{\ast}(t)\delta z_1-{{\delta z}_{1}}p
\end{equation}
\begin{equation} \label{6var3}
{{{\ddot{\delta z}}}_{2}}+{{\delta z}_{2}}={{T}_{2}}+\alpha \left( {{\delta z}_{1}}-{{\delta z}_{2}} \right) 
\end{equation}
\begin{equation} \label{6var4}
{{{\dot{\delta T}}}_{2}}+{{\delta T}_{2}}=2z_2^{\ast}(t) \delta z_{2}-{{\delta z}_{2}}p. 
\end{equation}
\end{subequations}
Unfortunately, these equations do not uncouple upon introducing the substitution Eq.~\eqref{transform}. This is because a strict OP mode is a not a solution of Eqs.~\eqref{coupled}. Hence, numerical simulation is resorted to for the solution of the above system. It yields that the perturbations remain bounded for $\alpha <0.88$ while they grow exponentially for $\alpha >0.88$. Numerical integration of the full system Eqs.~\eqref{coupled} yields a transition from stable to unstable at approximately $\alpha = 0.82$ which is in good agreement with the above prediction. 

\begin{figure}
	\centering	
	\includegraphics[width=0.5\textwidth]{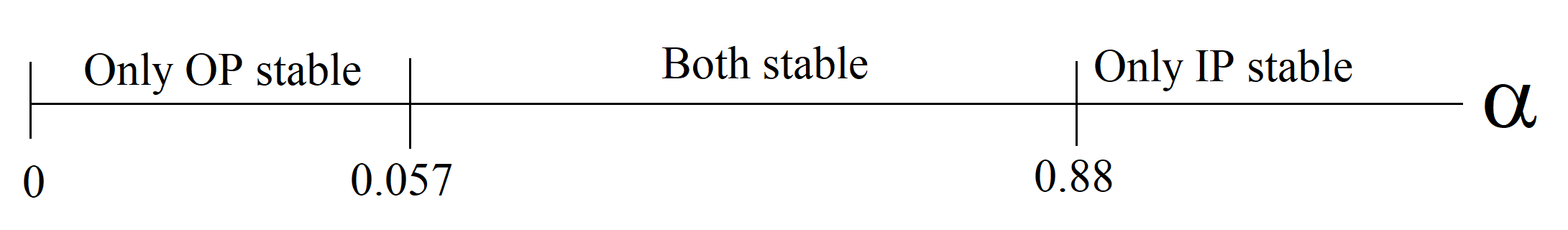}
	\caption{Stabilty of the IP and OP modes.}
	\label{fig:stabilty}
\end{figure}

Fig.~\ref{fig:stabilty} summarizes the foregoing sections by showing the stability of the IP and OP modes as a function of $\alpha$.

\section{Two variable expansion method}

The method of two variable expansion\cite{NayfehMook}, also known as  the method of multiple scales \cite{Nayfeh}, will now be used to study the coupled system. In this process, time is split into two variables, a regular time $\xi=t$ and a slow time $\eta = \varepsilon t$. The time derivative can be written as $\mathrm{d} /\mathrm{d}t = \partial / \partial \xi + \varepsilon \partial / \partial \eta$ and the second and higher derivatives may be calculated similarly. The small parameter $\varepsilon$ is introduced into Eq.~\eqref{coupled} in the following manner :
\begin{subequations} \label{twovar}
\begin{equation} \label{twovar1}
{{{\ddot{z}}}_{1}}+{{z}_{1}}=\varepsilon {{T}_{1}}+\varepsilon ^{2} \alpha \left( {{z}_{2}}-{{z}_{1}} \right)
\end{equation}
\begin{equation} \label{twovar2}
{{{\dot{T}}}_{1}}+{{T}_{1}}=z_{1}^{2}-\varepsilon {{z}_{1}}p 
\end{equation}
\begin{equation} \label{twovar3}
{{{\ddot{z}}}_{2}}+{{z}_{2}}=\varepsilon {{T}_{2}}+\varepsilon ^{2} \alpha \left( {{z}_{1}}-{{z}_{2}} \right) 
	\end{equation}
\begin{equation} \label{twovar4}
{{{\dot{T}}}_{2}}+{{T}_{2}}=z_{2}^{2}-\varepsilon {{z}_{2}}p 
\end{equation}
\end{subequations}
This scaling is motivated as follows : it is reasoned that the coupling, $\alpha$ is very weak ($O(\varepsilon^{2})$) but on account of the coupling, larger values of the static position $z_0$ can now be permitted. The following ansatz is attempted :
\begin{subequations} \label{ansatz}
\begin{equation} \label{ansatz1}
	 z_{1}=A(\eta )\cos \xi + B(\eta ) \sin \xi 
	\end{equation}
\begin{equation} \label{ansatz2}
	 z_{2}=C(\eta )\cos \xi + D(\eta ) \sin \xi \end{equation}
\end{subequations}
and additional dynamical variables are not taken for $T_1$ and $T_2$ since they are exponentially decaying at largest order. Substituting Eqs.~\eqref{ansatz} into Eqs.~\eqref{twovar} and removing resonant terms leads to the trivial slow flow $A'=B'=C'=D'=0$ where prime denotes $\mathrm{d}/\mathrm{d} \eta$. This indicates that the method has to be carried out to one further order. Defining a super-slow time $\zeta = \varepsilon ^{2} t$, $A, B, C, D$ are now taken to be functions of $\zeta$. Removal of resonance terms from the resulting equations now gives the following slow flow
\begin{subequations} \label{sloflo}
\begin{equation*} 
\frac{\text{d}A}{\text{d}\zeta }=(-1/120)\bigg[ 60\alpha D+31{{B}^{3}}+27A{{B}^{2}}+31A^2B
\end{equation*}
\begin{equation} \label{sloflo1}
-\left(30p+60\alpha  \right)B+27{{A}^{3}}-30pA \bigg] 
	\end{equation}
\begin{equation*} 
	\frac{\text{d}B}{\text{d}\zeta }=(1/120) \bigg[ 60\alpha C-27{{B}^{3}}+31A{{B}^{2}}+\left( 30p-27{{A}^{2}} \right)B
	\end{equation*}
	\begin{equation} \label{sloflo2}-\left( 30p+60\alpha  \right)A \bigg] 
		\end{equation}
\begin{equation*} 
	 \frac{\text{d}C}{\text{d}\zeta }	 =(-1/120) \bigg[ 60\alpha B+31{{D}^{3}}+27C{{D}^{2}}+ 31{{C}^{2}}D
\end{equation*} 
\begin{equation} \label{sloflo3}	 
-\left(30p+60\alpha  \right)D+27{{C}^{3}}-30pC \bigg]
		\end{equation}
\begin{equation*} 
	 \frac{\text{d}D}{\text{d}\zeta }=(1/120) \bigg[ 60\alpha A-27{{D}^{3}}+31C{{D}^{2}}+\left( 30p-27{{C}^{2}} \right)D
	\end{equation*}
	
\begin{equation} \label{sloflo4}
-\left( 30p+60\alpha  \right)C \bigg]
	\end{equation}
\end{subequations}

For increased convenience one can express the above system in terms of polar coordinates i.e. $A = r_{1} \cos \theta _{1}$, $B = r_{1} \sin \theta _{1}$, $C = r_{2} \cos \theta _{2}$ and $D = r_{2} \sin \theta _{2}$. If one defines the phase difference $\varphi = \theta _{2} - \theta_{1}$ then the following system for $r_{1}$, $r_{2}$ and $\varphi$ is obtained from Eq.~\eqref{sloflo}:
\begin{subequations} \label{polar}
\begin{equation} \label{polar1}
	 \frac{\text{d}{{r}_{1}}}{\text{d}\zeta }=\frac{p{{r}_{1}}}{4}-\frac{9r_{1}^{3}}{40}-\frac{\alpha }{2}{{r}_{2}}\sin \varphi 
	 	\end{equation}
\begin{equation} \label{polar2}
	 \frac{\text{d}{{r}_{2}}}{\text{d}\zeta }=\frac{p{{r}_{2}}}{4}-\frac{9r_{2}^{3}}{40}+\frac{\alpha }{2}{{r}_{1}}\sin \varphi 
	 	\end{equation}
\begin{equation} \label{polar2}
	\frac{\text{d}\varphi }{\text{d}\zeta }=\frac{31}{120}\left( r_{2}^{2}-r_{1}^{2} \right)+\frac{\alpha }{2}\cos \varphi \left( \frac{{{r}_{1}}}{{{r}_{2}}}-\frac{{{r}_{2}}}{{{r}_{1}}} \right)
	\end{equation}
\end{subequations}

Note that Eqs.~\eqref{polar} exhibit a symmetry: they are invariant under the transformation
\begin{equation}
r_1 \longrightarrow  r_2,~~~~r_2 \longrightarrow  r_1,~~~~\varphi \longrightarrow -\varphi
\label{sym}
\end{equation}

By inspection it can be seen that there are fixed points when $r_{1}=r_{2}$ and $\sin \varphi =0$ i.e. $\varphi = 0$ or $\varphi = \pi$. The former corresponds to the IP mode while the latter is the OP mode. A calculation for $r$ (either 1 or 2) at the fixed point yields $\frac{\sqrt{10p}}{3}$, which is same as that obtained from the Lindstedt analysis of Section \ref{section2}. To obtain the stability of the IP mode, the Jacobian is constructed for Eq.~\eqref{polar},
\begin{equation} \label{Jacob}
	\left[ \begin{matrix}
		-\frac{p}{2} & 0 & -\frac{\alpha \sqrt{10p}}{6}  \\
		0 & -\frac{p}{2} & \frac{\alpha \sqrt{10p}}{6}  \\
		\frac{-31\sqrt{10}{{p}^{3/2}}+54\alpha \sqrt{10p}}{180p} & \frac{31\sqrt{10}{{p}^{3/2}}-54\alpha \sqrt{10p}}{180p} & 0  \\
	\end{matrix} \right] 
\end{equation}
For $p=0.1$, the critical value of $\alpha$ is  31/540 = 0.057, where the IP mode is stable above this critical value and unstable below it. This is in good agreement with the variational equation and the results of numerical integration. Construction of the Jacobian for the OP mode shows that it remains stable at all values of $\alpha$. Thus, the two-variable method has yielded correctly the stability transition of the IP mode. However it cannot yield the transition for the OP mode, at this level of the perturbation theory.

\begin{figure}
	\centering
	\includegraphics[width=0.5\textwidth]{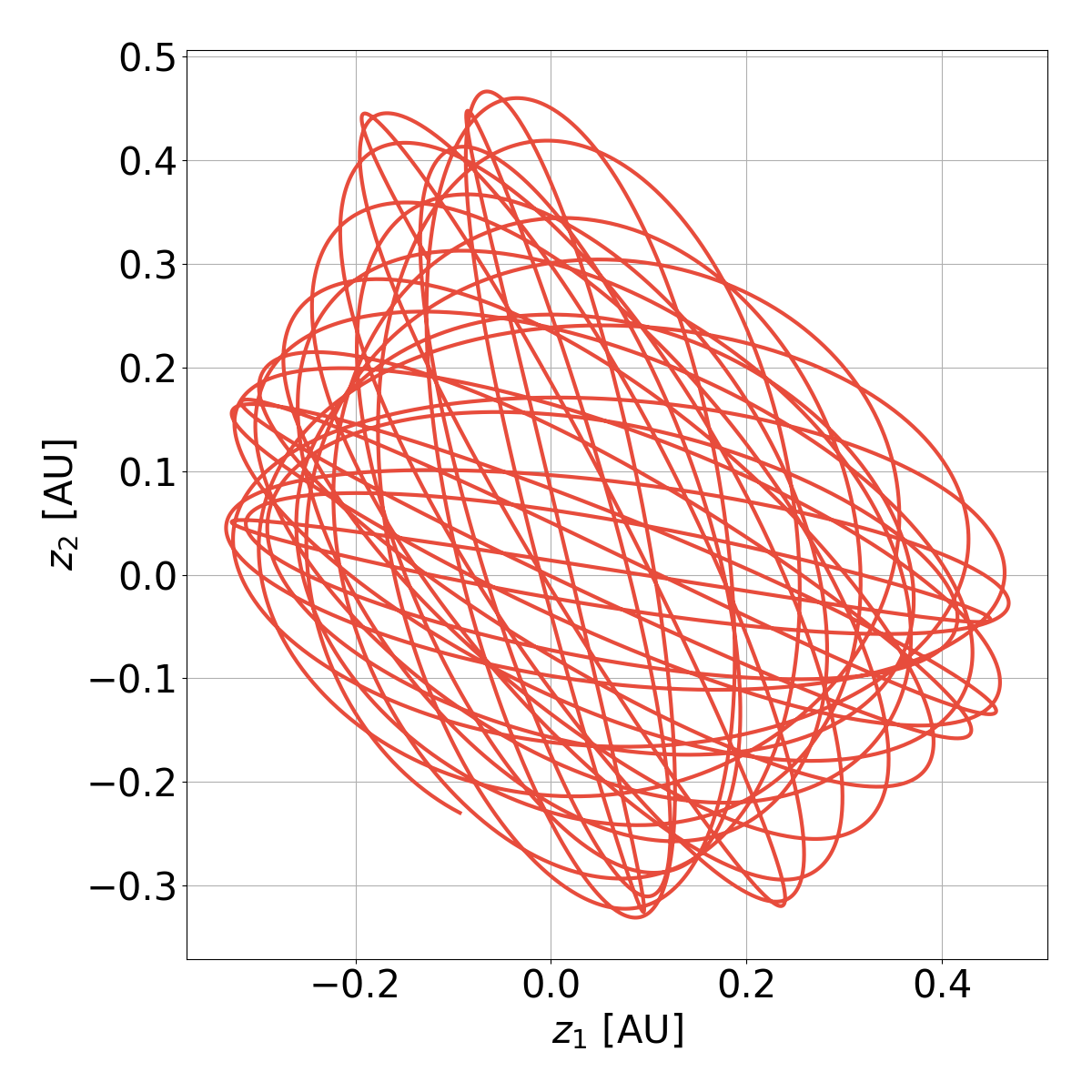}
	\caption{The separatrix exhibiting quasi-periodic motion in the phase plane. This plot is based on the numerical integration of the original system equations.}
	\label{fig: separatrix}
\end{figure}

\begin{figure}
\centering
	\includegraphics[width=0.5\textwidth]{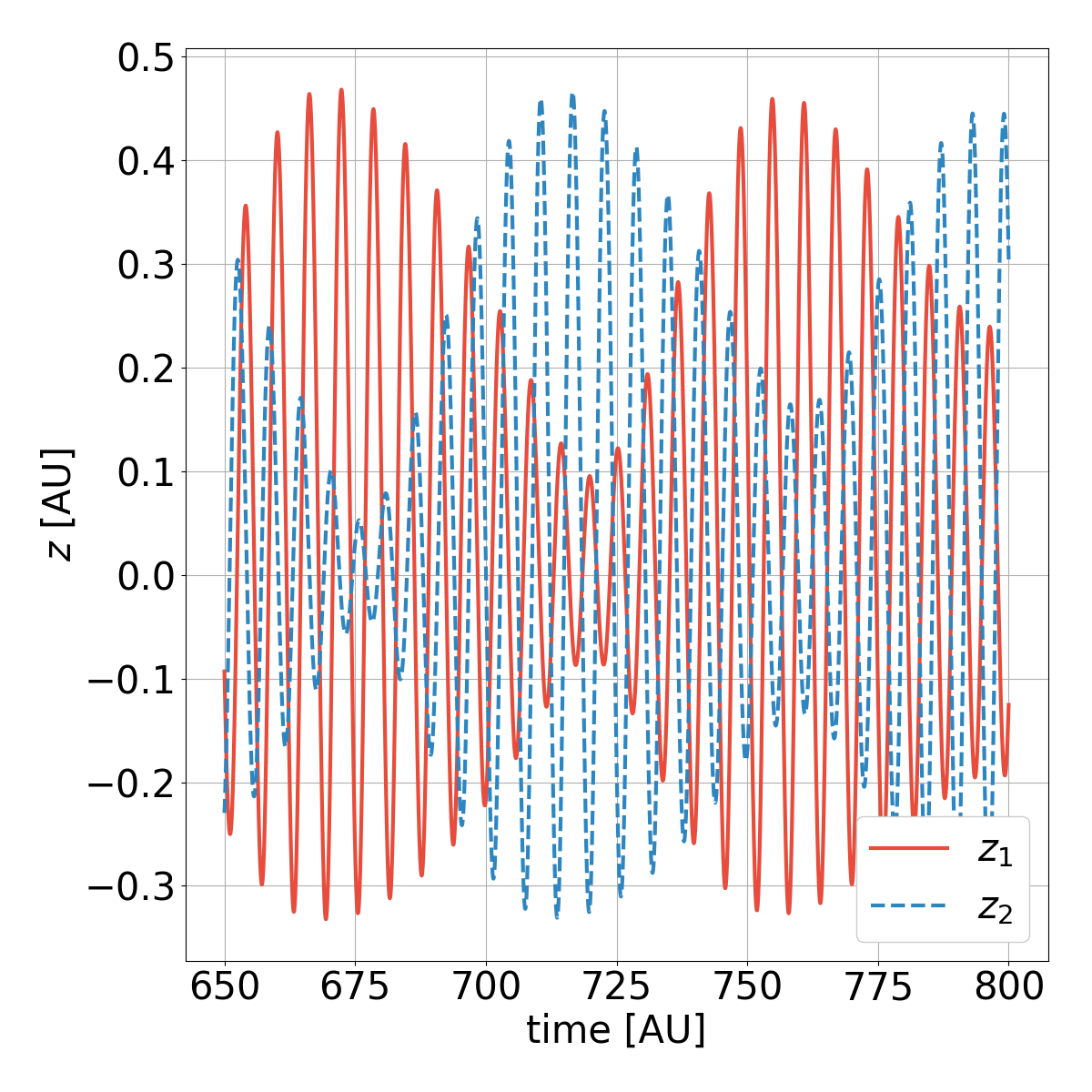}
		\caption{Time trace of the displacement variables at the separatrix. This plot is based on the the numerical resolution of the original system equations.}
		\label{fig: timetrace}
\end{figure}

Recall that we have shown that there is a range of $\alpha$ values for which both the IP and OP modes are stable, Fig.~\ref{fig:stabilty}. It is noted that these modes are limit cycles in the actual dynamical system but fixed points in the slow flow.  These slow flow equilibria are separated by an unstable slow flow limit cycle which we shall refer to as a separatrix.  Although it is unstable, one may nevertheless see what the separatrix looks like by choosing initial conditions to (approximately) lie on the basin boundary between the two equilibria.  Moving from the three-dimensional slow flow space to the 6 dimensional space of Eqs.~(\ref{coupled}), with initial conditions $\{z_1(0), z_2(0), \dot{z}_1(0), \dot{z}_2(0), T_1(0), T_2(0)\} = \{0.1, 0.00209816,$ \\$ 0, 0, 0, 0\}$ which are very close to the threshold, the separatrix appears as a quasi periodic motion, seen in Figs.~\ref{fig: separatrix} and \ref{fig: timetrace}. Such a situation was already encountered in Storti and Rand \cite{Storti}.\\

\begin{figure}
	\centering	
	\includegraphics[width=0.5\textwidth]{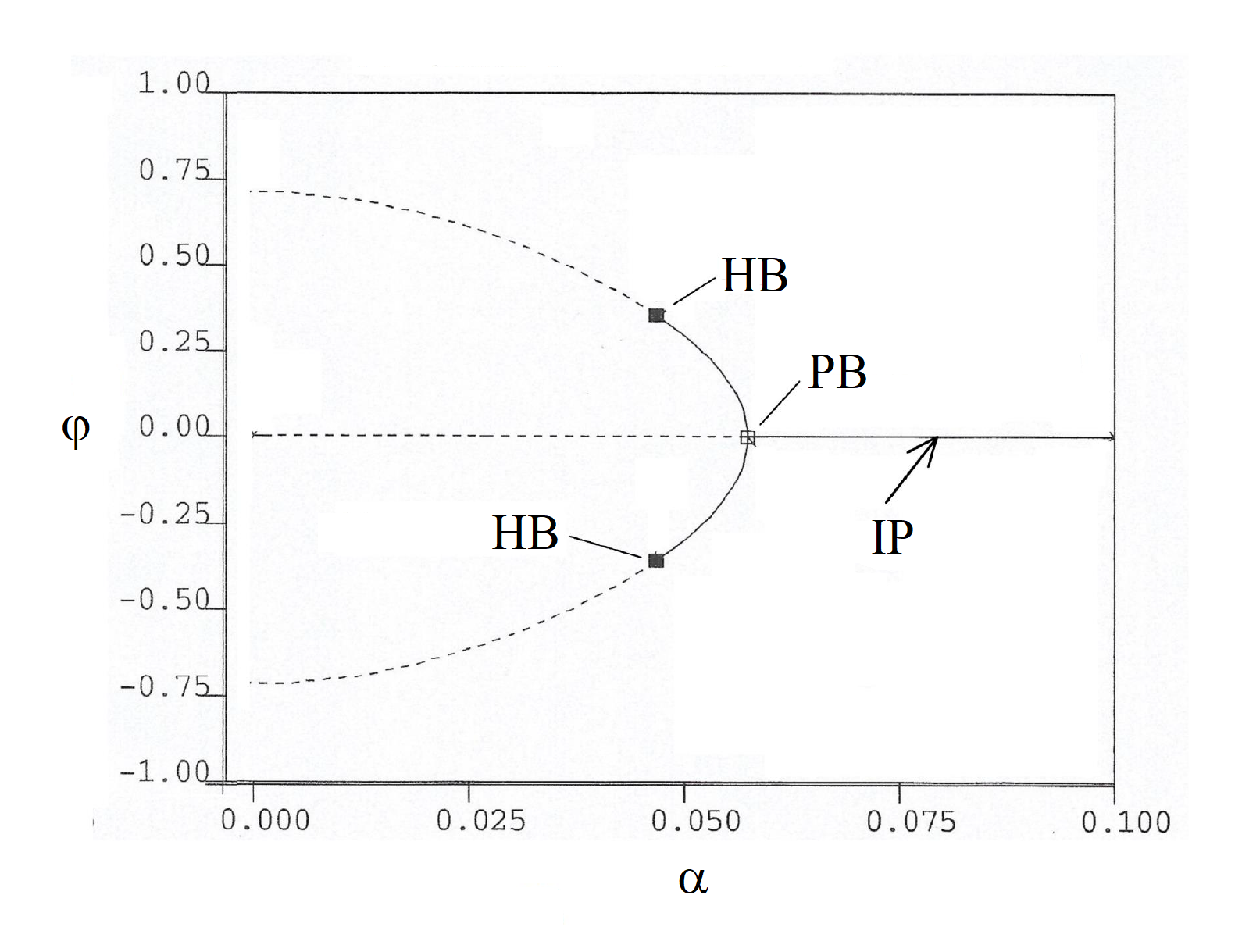}
	\caption{Bifurcation diagram from the slow flow equations, obtained from AUTO software. Here IP = In-phase mode, PB = Pitchfork bifurcation and HB = Hopf bifurcation.}
	\label{fig: AUTO}
\end{figure}

AUTO \cite{AUTOcite} is an analytic continuation software package which we will use to plot the bifurcation diagram of the slow flow Eqs.\eqref{polar}. The result is shown in Fig. \ref{fig: AUTO} where slow flow fixed points are displayed in the $\varphi$ vs. $\alpha$ plane. It can be seen that the IP mode is stable up to $\alpha = 0.057$; thereafter it cedes stability to a pair of slow flow fixed points which are born in a pitchfork bifurcation.

\begin{figure}
	\centering
	\includegraphics[width=0.5\textwidth]{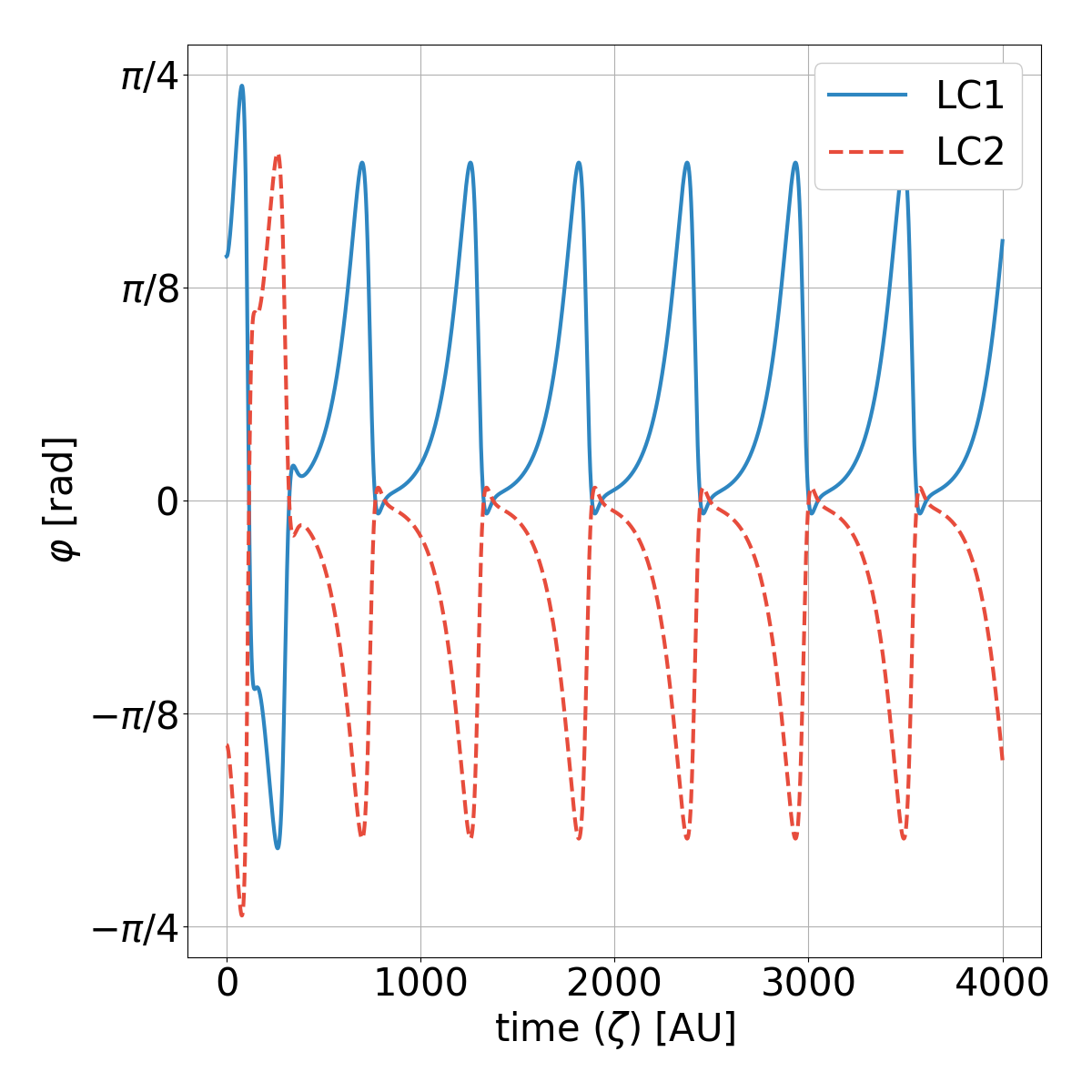}
	\caption{Mirror symmetric limit cycles, LC1 and LC2, at $\alpha=0.0437$ prior to the homoclinic bifurcation which occurs at $\alpha =0.0436 $, as the parameter $\alpha$ is decreased.}
	\label{fig: superpose}
\end{figure}

\begin{figure}
	\centering
	\includegraphics[width=0.5\textwidth]{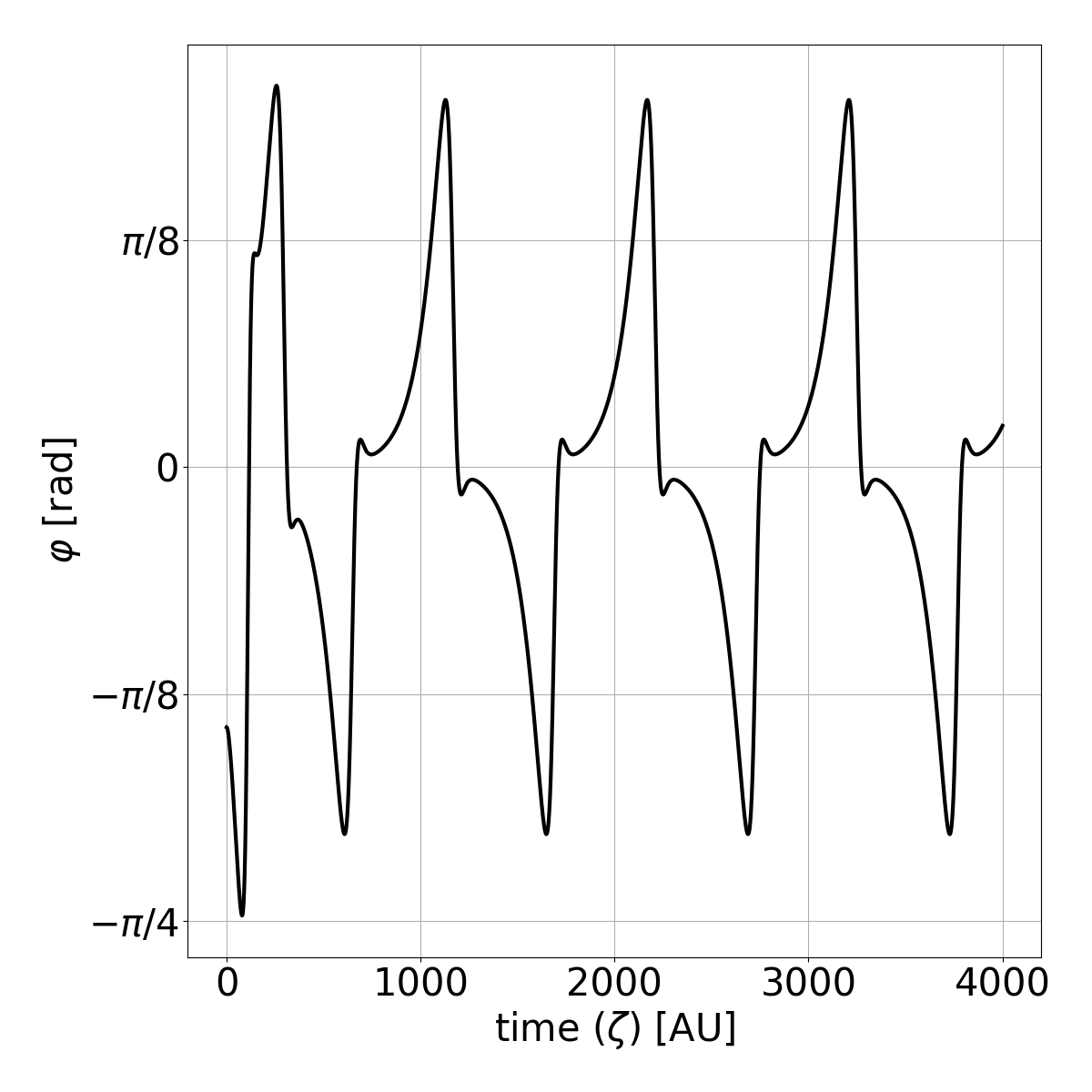}
	\caption{Single limit cycle at $\alpha=0.0435$ after the homoclinic bifurcation which occurs at $\alpha =0.0436 $, as the parameter $\alpha$ is decreased.}
	\label{fig: solo}
\end{figure}

Since the slow flow system Eq.~\eqref{polar} possesses the symmetry Eq.~\eqref{sym}, all dynamical quantities either have this symmetry or are part of a pair of reflected twins. 
 AUTO shows that these slow flow fixed points lose stability in a Hopf bifurcation at approximately $\alpha = 0.0468$; the resulting symmetric slow flow limit cycles are stable. 
Numerical integration shows that these limit cycles increase in amplitude and deform in shape as $\alpha$ is decreased further, up to a further bifurcation which occurs when $\alpha$ decreases through approximately 0.0436, though this is not shown in Fig. \ref{fig: AUTO}.  In this case there is
a homoclinic bifurcation in which the two asymmetric slow flow limit cycles join to become a single slow flow limit cycle  which exhibits the symmetry of Eq.~\eqref{sym}. In Fig.~\ref{fig: superpose}, one sees the two limit cycles obtained by numerical integration of the slow flow with parameter value $\alpha = 0.0437$ and initial conditions $r_1=r_2=1/3$ and $\varphi = +0.45$ and $\varphi = -0.45$ respectively. In Fig.~\ref{fig: solo}, one can see the single symmetric limit cycle obtained  at $\alpha = 0.0435$ and starting from either of the two initial conditions used in obtaining the mirror symmetric limit cycles, earlier.

 Another bifurcation occurs when $\alpha$ decreases through approximately 0.0415, in which the unstable separatrix limit cycle (Fig.~\ref{fig: separatrix}) merges with the symmetric slow flow stable limit cycle which was created in the homoclinic bifurcation.  For values of $\alpha$ less than approximately  0.0415, the OP mode is the only stable motion.

\begin{figure}
	\centering
	\includegraphics[width=0.5\textwidth]{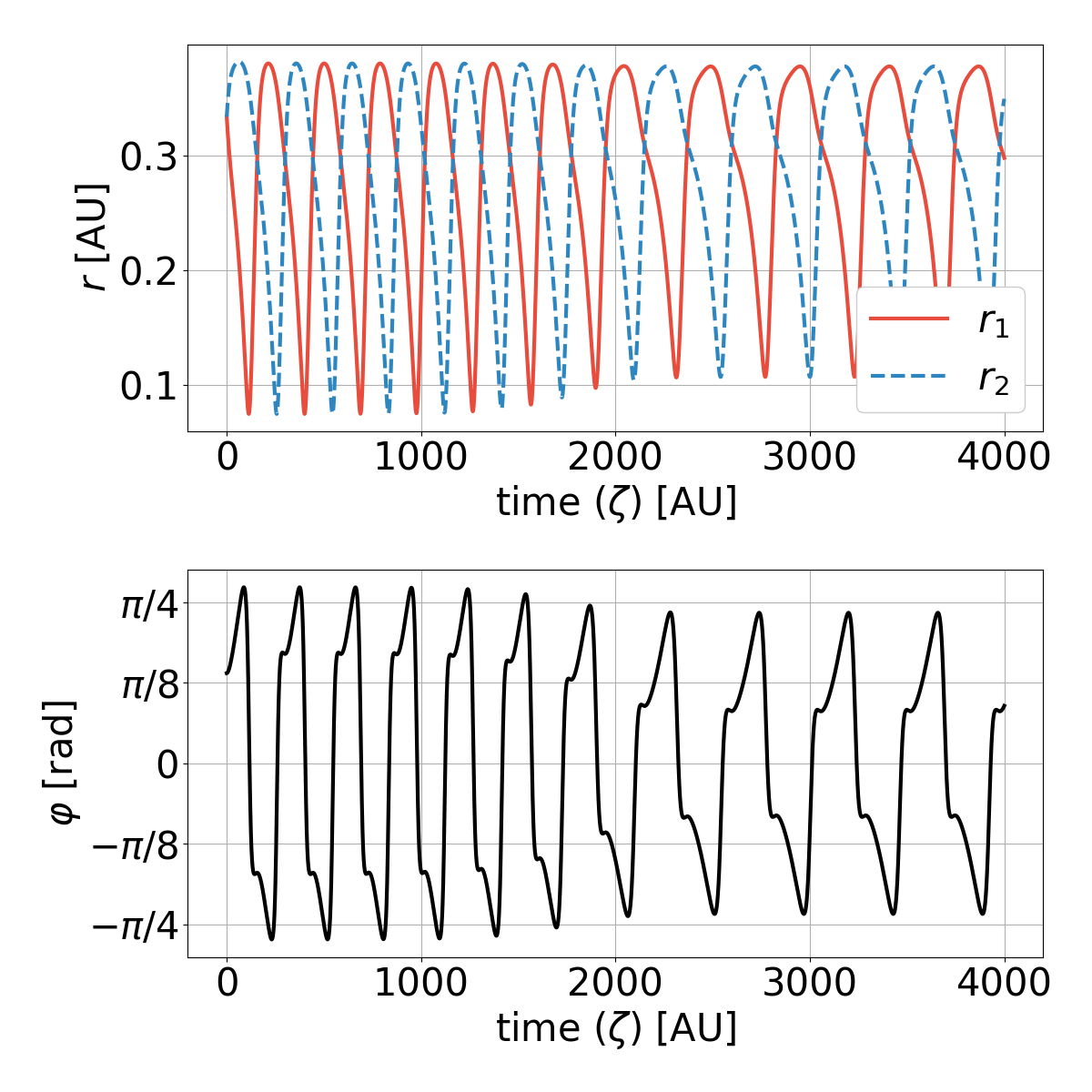}
	\caption{Numerical integration of the slow flow showing the system starting on the separatrix at $\alpha = 0.042$ and going over to the steady state limit cycle.}
	\label{fig: fig11}
\end{figure}

\begin{figure}
\centering
	\includegraphics[width=0.5\textwidth]{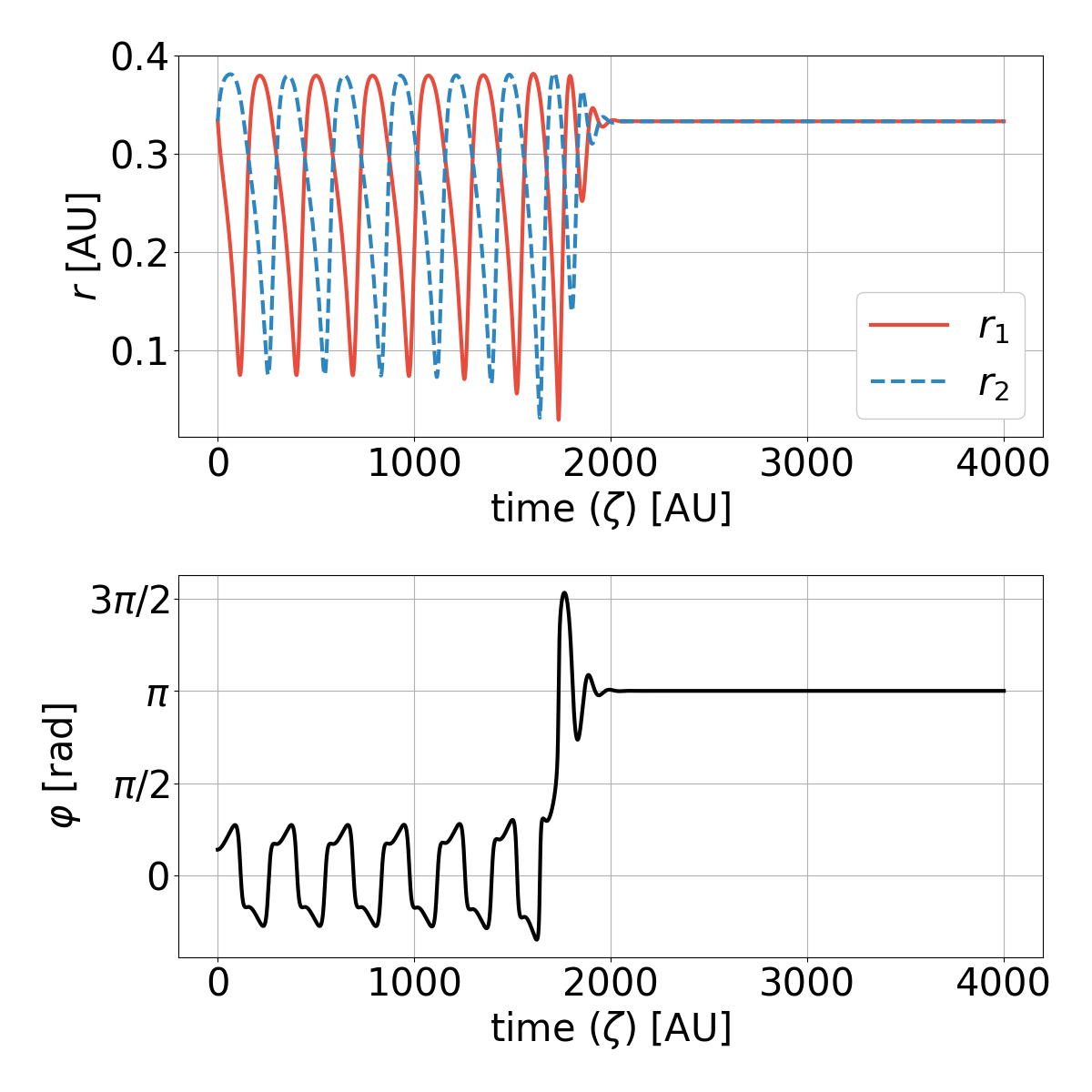}
		\caption{Slow flow system starting on the separatrix at $\alpha = 0.042$  and going over to the OP mode.}
		\label{fig: fig12}
\end{figure}

\begin{figure}
\centering
	\includegraphics[width=0.5\textwidth]{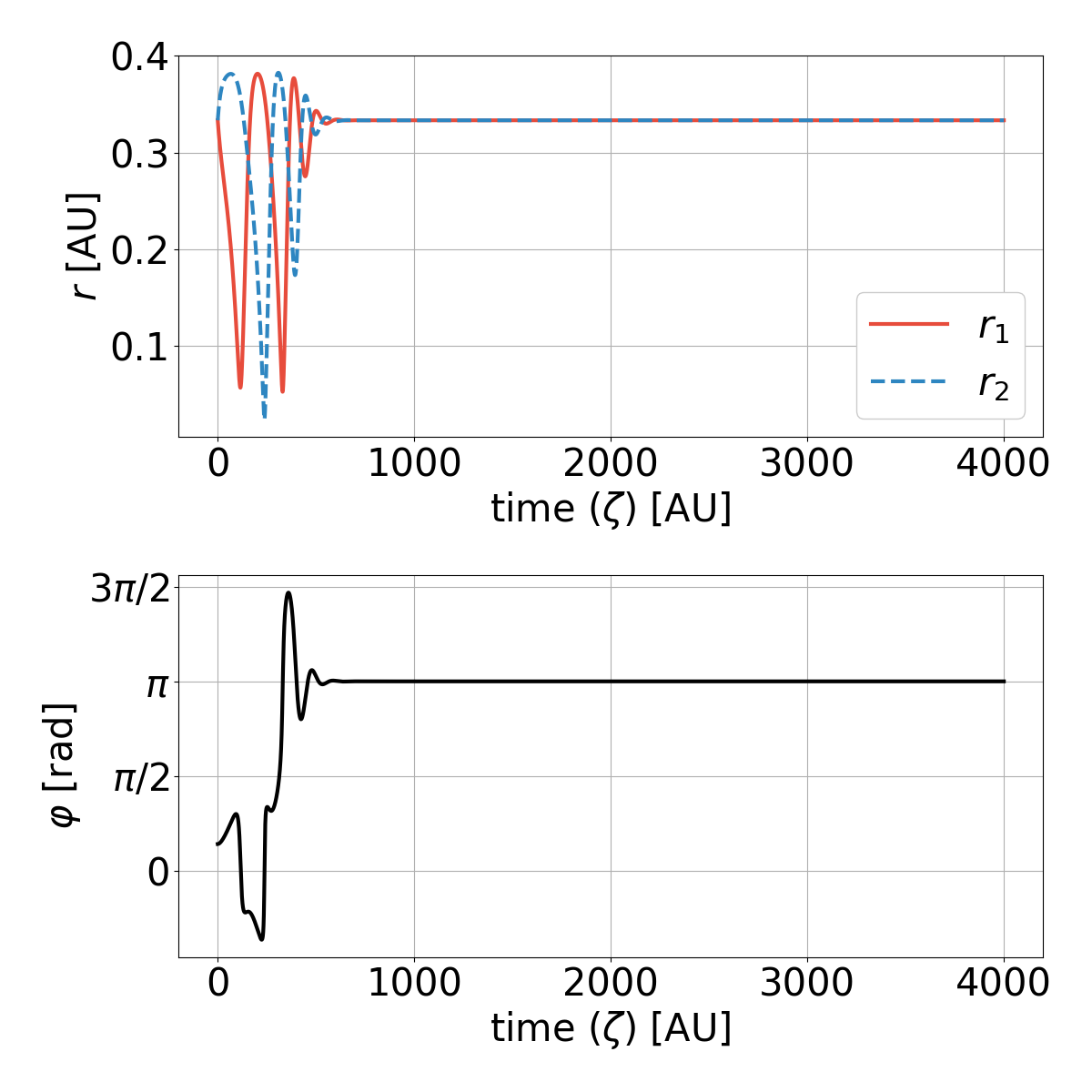}
		\caption{At $\alpha = 0.041$, all initial conditions lead the slow flow system to go to the OP mode.}
		\label{fig: fig13}
\end{figure}

To visualize the bifurcation at $\alpha=0.0415$, the results of simulation are shown with IC picked on the crack between attainment of limit cycle and transition to OP mode. Fig.~\ref{fig: fig11} presents the results of numerical integration of the slow flow with parameter value $\alpha = 0.042$ and ICs $r_1 = r_2 = 1/3$ and $\varphi = 0.4403$ while in Fig.~\ref{fig: fig12} we change the IC $\varphi$ to 0.4404. At this $\alpha$ value, there are two stable objects; the symmetric limit cycle and the fixed point corresponding to the OP mode. In these simulations it can be seen that the system remains on the separatrix for a short time before it progresses to its asymptotic state. In Fig.~\ref{fig: fig13}, the value of $\alpha$ is changed to 0.041. Starting at the initial conditions corresponding to  Fig.~\ref{fig: fig11}, the system goes to the OP mode.

\section{Conclusions}

Thus, in this article a study of a simplified model of coupled MEMS oscillators is performed. The most interesting feature of the system is that the IP and the OP modes are both stable over a considerable range of parameter values. For sufficiently low $\alpha$ $(<0.04)$ only the OP mode is stable. At intermediate $\alpha$, both the IP and the OP modes coexist stably. At high $\alpha$ $(>0.88)$ only the IP mode is stable. This coexistent stability of the two modes is reminiscent of what was found in Reference \cite{Storti} for two coupled van der Pol oscillators. Table \ref{tab:1} shows the bifurcations that occur in this system as the value of $\alpha$ is varied. 

\begin{table}
\caption{Stability of the system to different values of $\alpha$.} 
\label{tab:1}       
\begin{tabular}{ll}
\hline\noalign{\smallskip}
Value of $\alpha$ & Stability \\
\noalign{\smallskip}\hline\noalign{\smallskip}
$>0.9$ & only IP \\
$0.88$   & OP becomes stable \\
$0.057$ & IP loses stability in \\
&pitchfork to modified IP \\
$0.0468$ & modified IP loses stability \\
&in Hopf to quasiperiodic \\
$0.0436$ & two LCs coalesce in homoclinic \\
$0.042$ & stable LC collides with separatrix\\
& and vanishes \\
$<0.042$ & only OP \\
\noalign{\smallskip}\hline
\end{tabular}
\end{table}

As regards the various methods adopted, it is noted that the two variable method yielded the most accurate results in the regions where it was effective. However, it is ineffective in predicting the stability of OP mode. It is also an interesting fact that this is one system which is driven by a third order equation, and its stability analysis led us to a third order Mathieu-like equation. It is likely that these studies can be extended to incorporate other third order parametrically excited systems and MEMS systems with detuned oscillators in the future.

\section{Acknowledgements}

This material is based upon work supported by the National Science Foundation under grant number CMMI-1634664. The authors wish to thank Professor John Guckenheimer for advising them on the bifurcations involved in this paper.

\section{Compliance with ethical standards}
There is no conflict of interest. Funding has been received from NSF as acknowledged above. The entire research presented here is the authors' own. No part of this Article has been reproduced from other Articles or is under consideration elsewhere. Research on human and animal subjects was not necessary for this project. All authors consent to submission of this Article in its present form.

\bibliographystyle{spmpsci}      
\bibliography{Biblio}   

\end{document}